\documentclass[reqno,12pt,a4paper]{amsart}

\usepackage{mathrsfs}
\usepackage{amssymb}
\usepackage{amsfonts}
\usepackage{latexsym}
\usepackage{amsthm}

\usepackage{graphicx}
\def\lb{\label}

\newcommand{\er}[1]{\textrm{(\ref{#1})}}

\begin{document}


\renewcommand{\theequation}{\arabic{section}.\arabic{equation}}
\theoremstyle{plain}
\newtheorem{theorem}{\bf Theorem}[section]
\newtheorem{lemma}[theorem]{\bf Lemma}
\newtheorem{corollary}[theorem]{\bf Corollary}
\newtheorem{proposition}[theorem]{\bf Proposition}
\newtheorem{definition}[theorem]{\bf Definition}
\newtheorem{remark}[theorem]{\it Remark}

\def\a{\alpha}  \def\cA{{\mathcal A}}     \def\bA{{\bf A}}  \def\mA{{\mathscr A}}
\def\b{\beta}   \def\cB{{\mathcal B}}     \def\bB{{\bf B}}  \def\mB{{\mathscr B}}
\def\g{\gamma}  \def\cC{{\mathcal C}}     \def\bC{{\bf C}}  \def\mC{{\mathscr C}}
\def\G{\Gamma}  \def\cD{{\mathcal D}}     \def\bD{{\bf D}}  \def\mD{{\mathscr D}}
\def\d{\delta}  \def\cE{{\mathcal E}}     \def\bE{{\bf E}}  \def\mE{{\mathscr E}}
\def\D{\Delta}  \def\cF{{\mathcal F}}     \def\bF{{\bf F}}  \def\mF{{\mathscr F}}
\def\c{\chi}    \def\cG{{\mathcal G}}     \def\bG{{\bf G}}  \def\mG{{\mathscr G}}
\def\z{\zeta}   \def\cH{{\mathcal H}}     \def\bH{{\bf H}}  \def\mH{{\mathscr H}}
\def\e{\eta}    \def\cI{{\mathcal I}}     \def\bI{{\bf I}}  \def\mI{{\mathscr I}}
\def\p{\psi}    \def\cJ{{\mathcal J}}     \def\bJ{{\bf J}}  \def\mJ{{\mathscr J}}
\def\vT{\Theta} \def\cK{{\mathcal K}}     \def\bK{{\bf K}}  \def\mK{{\mathscr K}}
\def\k{\kappa}  \def\cL{{\mathcal L}}     \def\bL{{\bf L}}  \def\mL{{\mathscr L}}
\def\l{\lambda} \def\cM{{\mathcal M}}     \def\bM{{\bf M}}  \def\mM{{\mathscr M}}
\def\L{\Lambda} \def\cN{{\mathcal N}}     \def\bN{{\bf N}}  \def\mN{{\mathscr N}}
\def\m{\mu}     \def\cO{{\mathcal O}}     \def\bO{{\bf O}}  \def\mO{{\mathscr O}}
\def\n{\nu}     \def\cP{{\mathcal P}}     \def\bP{{\bf P}}  \def\mP{{\mathscr P}}
\def\r{\rho}    \def\cQ{{\mathcal Q}}     \def\bQ{{\bf Q}}  \def\mQ{{\mathscr Q}}
\def\s{\sigma}  \def\cR{{\mathcal R}}     \def\bR{{\bf R}}  \def\mR{{\mathscr R}}
                \def\cS{{\mathcal S}}     \def\bS{{\bf S}}  \def\mS{{\mathscr S}}
\def\t{\tau}    \def\cT{{\mathcal T}}     \def\bT{{\bf T}}  \def\mT{{\mathscr T}}
\def\f{\phi}    \def\cU{{\mathcal U}}     \def\bU{{\bf U}}  \def\mU{{\mathscr U}}
\def\F{\Phi}    \def\cV{{\mathcal V}}     \def\bV{{\bf V}}  \def\mV{{\mathscr V}}
\def\P{\Psi}    \def\cW{{\mathcal W}}     \def\bW{{\bf W}}  \def\mW{{\mathscr W}}
\def\o{\omega}  \def\cX{{\mathcal X}}     \def\bX{{\bf X}}  \def\mX{{\mathscr X}}
\def\x{\xi}     \def\cY{{\mathcal Y}}     \def\bY{{\bf Y}}  \def\mY{{\mathscr Y}}
\def\X{\Xi}     \def\cZ{{\mathcal Z}}     \def\bZ{{\bf Z}}  \def\mZ{{\mathscr Z}}
\def\O{\Omega}

\newcommand{\gA}{\mathfrak{A}}
\newcommand{\gB}{\mathfrak{B}}
\newcommand{\gC}{\mathfrak{C}}
\newcommand{\gD}{\mathfrak{D}}
\newcommand{\gE}{\mathfrak{E}}
\newcommand{\gF}{\mathfrak{F}}
\newcommand{\gG}{\mathfrak{G}}
\newcommand{\gH}{\mathfrak{H}}
\newcommand{\gI}{\mathfrak{I}}
\newcommand{\gJ}{\mathfrak{J}}
\newcommand{\gK}{\mathfrak{K}}
\newcommand{\gL}{\mathfrak{L}}
\newcommand{\gM}{\mathfrak{M}}
\newcommand{\gN}{\mathfrak{N}}
\newcommand{\gO}{\mathfrak{O}}
\newcommand{\gP}{\mathfrak{P}}
\newcommand{\gQ}{\mathfrak{Q}}
\newcommand{\gR}{\mathfrak{R}}
\newcommand{\gS}{\mathfrak{S}}
\newcommand{\gT}{\mathfrak{T}}
\newcommand{\gU}{\mathfrak{U}}
\newcommand{\gV}{\mathfrak{V}}
\newcommand{\gW}{\mathfrak{W}}
\newcommand{\gX}{\mathfrak{X}}
\newcommand{\gY}{\mathfrak{Y}}
\newcommand{\gZ}{\mathfrak{Z}}

\def\ve{\varepsilon} \def\vt{\vartheta} \def\vp{\varphi}  \def\vk{\varkappa}

\def\Z{{\mathbb Z}} \def\R{{\mathbb R}} \def\C{{\mathbb C}}  \def\K{{\mathbb K}}
\def\T{{\mathbb T}} \def\N{{\mathbb N}} \def\dD{{\mathbb D}} \def\S{{\mathbb S}}
\def\B{{\mathbb B}}


\def\la{\leftarrow}              \def\ra{\rightarrow}     \def\Ra{\Rightarrow}
\def\ua{\uparrow}                \def\da{\downarrow}
\def\lra{\leftrightarrow}        \def\Lra{\Leftrightarrow}
\newcommand{\abs}[1]{\lvert#1\rvert}
\newcommand{\br}[1]{\left(#1\right)}

\def\lan{\langle} \def\ran{\rangle}


\def\lt{\biggl}                  \def\rt{\biggr}
\def\ol{\overline}               \def\wt{\widetilde}
\def\no{\noindent}


\let\ge\geqslant                 \let\le\leqslant
\def\lan{\langle}                \def\ran{\rangle}
\def\/{\over}                    \def\iy{\infty}
\def\sm{\setminus}               \def\es{\emptyset}
\def\ss{\subset}                 \def\ts{\times}
\def\pa{\partial}                \def\os{\oplus}
\def\om{\ominus}                 \def\ev{\equiv}
\def\iint{\int\!\!\!\int}        \def\iintt{\mathop{\int\!\!\int\!\!\dots\!\!\int}\limits}
\def\el2{\ell^{\,2}}             \def\1{1\!\!1}
\def\sh{\sharp}
\def\wh{\widehat}
\def\bs{\backslash}
\def\na{\nabla}

\def\sh{\mathop{\mathrm{sh}}\nolimits}
\def\all{\mathop{\mathrm{all}}\nolimits}
\def\Area{\mathop{\mathrm{Area}}\nolimits}
\def\arg{\mathop{\mathrm{arg}}\nolimits}
\def\const{\mathop{\mathrm{const}}\nolimits}
\def\det{\mathop{\mathrm{det}}\nolimits}
\def\diag{\mathop{\mathrm{diag}}\nolimits}
\def\diam{\mathop{\mathrm{diam}}\nolimits}
\def\dim{\mathop{\mathrm{dim}}\nolimits}
\def\dist{\mathop{\mathrm{dist}}\nolimits}
\def\Im{\mathop{\mathrm{Im}}\nolimits}
\def\Iso{\mathop{\mathrm{Iso}}\nolimits}
\def\Ker{\mathop{\mathrm{Ker}}\nolimits}
\def\Lip{\mathop{\mathrm{Lip}}\nolimits}
\def\rank{\mathop{\mathrm{rank}}\limits}
\def\Ran{\mathop{\mathrm{Ran}}\nolimits}
\def\Re{\mathop{\mathrm{Re}}\nolimits}
\def\Res{\mathop{\mathrm{Res}}\nolimits}
\def\res{\mathop{\mathrm{res}}\limits}
\def\sign{\mathop{\mathrm{sign}}\nolimits}
\def\span{\mathop{\mathrm{span}}\nolimits}
\def\supp{\mathop{\mathrm{supp}}\nolimits}
\def\Tr{\mathop{\mathrm{Tr}}\nolimits}
\def\BBox{\hspace{1mm}\vrule height6pt width5.5pt depth0pt \hspace{6pt}}
\def\where{\mathop{\mathrm{where}}\nolimits}
\def\as{\mathop{\mathrm{as}}\nolimits}
\def\Dom{\mathop{\mathrm{Dom}}\nolimits}


\newcommand\nh[2]{\widehat{#1}\vphantom{#1}^{(#2)}}
\def\dia{\diamond}

\def\Oplus{\bigoplus\nolimits}



\def\qqq{\qquad}
\def\qq{\quad}
\let\ge\geqslant
\let\le\leqslant
\let\geq\geqslant
\let\leq\leqslant
\newcommand{\ca}{\begin{cases}}
\newcommand{\ac}{\end{cases}}
\newcommand{\ma}{\begin{pmatrix}}
\newcommand{\am}{\end{pmatrix}}
\renewcommand{\[}{\begin{equation}}
\renewcommand{\]}{\end{equation}}
\def\eq{\begin{equation}}
\def\qe{\end{equation}}
\def\[{\begin{equation}}
\def\bu{\bullet}

\newcommand{\fr}{\frac}
\newcommand{\tf}{\tfrac}

\title[{Asymptotics of determinants of 4-th order operators at zero}]
{Asymptotics of determinants of 4-th order operators at zero}

\date{\today}
\author[Andrey Badanin]{Andrey Badanin}
\author[Evgeny Korotyaev]{Evgeny L. Korotyaev}
\address{Saint-Petersburg
State University, Universitetskaya nab. 7/9, St. Petersburg, 199034
Russia, an.badanin@gmail.com, a.badanin@spbu.ru,
korotyaev@gmail.com,  e.korotyaev@spbu.ru}


\begin{abstract}
\no We consider fourth order ordinary differential operators with
compactly supported  coefficients  on the half-line and on the line.
The Fredholm determinant for this operator is an analytic function in the
whole complex plane without zero.
We describe the determinant at zero.
We show that in the generic case it
has a pole of order 4 in the case of the line
and of order 1 in the case of the half-line.
\end{abstract}

\maketitle

\section {Introduction and main results}
\setcounter{equation}{0}

We consider fourth order operators
$$
H=H_0+V,\qq\where\qq V=2\pa p\pa +q,\qq \pa ={d\/dx},
$$
and $H_0$ is one of the following:
$$
\begin{aligned}
&\text{Case 1:}\qqq \pa^4 \text{ in }  L^2(\R_+),
\text{ with } y(0)=y''(0)=0
\text{ boundary conditions}.
\\
&\text{Case 2:}\qqq \pa^4 \text{ in } L^2(\R).
\end{aligned}
$$
We assume that the functions $p,q$ are compactly supported
and belong to the space of functions $\cH$, defined by
$$
\begin{aligned}
&\cH=\big\{f\in L^1(\R_+): \supp f\in[0,\g]\big\}\qqq \text{ in Case 1}
\\
&\cH=\big\{\ f\in L^1(\R)\ :\supp f\in[0,\g]\big\}\qqq \text{ in Case 2}
\end{aligned}
$$
for some $\g>0$. We consider the operator $H_0$ as unperturbed and
the perturbed operator $H$ is defined on the form domain $
\Dom_{fd}(H)=\Dom_{fd}(H_0). $ The operator $H$ has purely
absolutely continuous spectrum $[0,\iy)$ plus a finite number of
simple eigenvalues.

We  rewrite $V$ in the form
\[
\lb{defVj}
V=V_1V_2,\qqq V_1=(\pa |2p|^{1\/2}, |q|^{1\/2}),\qqq
 V_2=\ma
(2p)^{1\/2}\pa
\\ q^{1\/2}\am,
\]
and we set
\[
\lb{Y0}
 R_0(k)=(H_0-k^4)^{-1},\qqq
k \in \K_1=\Big\{k\in\C: \arg k\in \Big(0,{\pi\/2}\Big)\Big\}.
\]
Each operator $V_2 R_0(k)V_1,k\in \K_1,$ is trace class and thus we can
define a Fredholm determinant:
\[
\label{a.2}
D(k)=\det \big(I +V_2 R_0(k)V_1\big),\qqq k\in \K_1,
\]
similar to second order operators.
The function $D$ is analytic in
$\K_1$ and has an analytic extension from $\K_1$ into the whole
complex plane without zero.
Note that $k\in\ol\K_1\sm\{0\}$ is a zero of
the determinant $D$ iff $\l=k^4\in \R\sm\{0\}$ is an eigenvalue of
the operator $H$. We define the resonances as zeros of the Fredholm
determinant in $\C\sm \ol \K_1$.
A main goal of the present paper is to describe the determinant
at the point zero.

Let $p,q\in\cH$. Consider the equation
\[
\lb{ie} y''''+2(py')'+qy=k^4 y,\qqq k\in\C,
\]
on the interval $[0,\g]$
in the class of function $y$ satisfying the conditions
$$
y^{[j]}\in AC([0,\g]),\qq j=0,1,2,3,
$$
where $AC([0,\g])$ is the class of absolutely continuous functions
on $[0,\g]$ and
$$
y^{[0]}=y,\qq y^{[1]}=y',\qq y^{[2]}=y'',\qq
y^{[3]}=y'''+2py'.
$$
Introduce the fundamental solutions
$\vp_j(x,k),j\in\N_4=\{1,2,3,4\}$, of equation \er{ie}
 satisfying the conditions
\[
\lb{iecond}
\vp_j^{[\ell -1]}(0,k)=\d_{j\ell },\qq
j,\ell\in\N_4.
\]
Introduce the $4\ts4$ matrix-valued function
$$
M(k)=(M_{j\ell}(k))_{j,\ell=1}^4
=\big(\vp_j^{[\ell-1]}(\g,k)\big)_{j,\ell=1}^4,\qqq k\in\C.
$$
Each function $\vp_j(x,\cdot),(j,x)\in\N_4\ts\R$, and $M$ is entire
in $\l=k^4$ and real at real $\l$.

Introduce the operators  $H_j=\pa^4+2p\pa p+q,j=1,2,3$,
in $L^2(0,\g)$ with the boundary
conditions
\[
\lb{bcDent}
\begin{aligned}
y(0)=y''(0)=y''(\g)=y^{[3]}(\g)=0\qqq\text{for}\qq H_1\\
y(0)=y''(0)=y'(\g)=y^{[3]}(\g)=0\qqq\text{for}\qq H_2,\\
y''(0)=y^{[3]}(0)=y''(\g)=y^{[3]}(\g)=0\qqq\text{for}\qq H_3,
\end{aligned}
\]
A spectrum $\s(H_j)$ of each operator $H_j,j=1,2,3$ is discrete.
Moreover,
\[
\lb{sps}
\begin{aligned}
\s(H_1)=\{\l=k^4\in\C:\D_{34 }(k)=0\},\\
\s(H_2)=\{\l=k^4\in\C:\D_{24 }(k)=0\},\\
\s(H_3)=\{\l=k^4\in\C:\F(k)=0\},
\end{aligned}
\]
where $\Phi(k), \D_{j\ell}(k),j,\ell\in\N_4$
are entire functions in $\l=k^4$, real at real $\l$,
and given by
\[
\lb{defD}
\D_{j\ell}=\det\ma M_{j 2}&M_{j 4}\\M_{\ell 2}&M_{\ell 4}\am,
\qqq
\F=\det\ma M_{31}& M_{32}\\ M_{41}& M_{42}\am.
\]
We prove the following main results.

\begin{theorem}
\lb{T2} Let $p,q\in\cH$ in Case 1. Then the function $kD(k)$ is entire
and  satisfies:
\[
\lb{asDhl}
D(k)=\fr{e^{(i-1)k\g}}{2k}\Big((1+i)\D_{34 }(0)
+2k\D_{24 }(0)
+(1-i)k^2\big(\D_{14 }(0)+\D_{23 }(0)\big)+O(k^{3})
\Big),
\]
as $k\to 0$ uniformly on $\arg k\in[0,2\pi]$.
Moreover,
\[
\lb{detne0}
\big|\D_{34 }(0)\big|+\big|\D_{24 }(0)\big|+\big|\D_{14 }(0)+\D_{23 }(0)\big|\ne 0,
\]
that is the function $k^{-2}D(k)$ has a pole of order $\ge 1$ at zero.
Furthermore,

(i) The function $D(k)$ is entire iff $\l=0$ is an eigenvalue of $H_1$,

(ii) The function $k^{-1}D(k)$ is entire iff $\l=0$ is an eigenvalue of both
$H_1$ and $H_2$.

\end{theorem}

\medskip

\begin{theorem}
\lb{ThDentl} Let $p,q\in\cH$ in Case 2. Then the function $k^4D(k)$ is
entire and  satisfies:
\[
\lb{asDat0l}
D(k)=-\fr{1}{8k^4}
\big(\F(0)+O(k)\big),
\]
as $k\to 0$ uniformly on $\arg k\in[0,2\pi]$. In particular, the
function $k^3D(k)$ is entire iff $\l=0$ is
an eigenvalue of $H_3$.

\end{theorem}

\no {\bf Remark.}
1) It is well known that the Fredholm determinant
for a Scr\"odinger operator on the half-line
with compactly supported potential
is an entire function, and the determinant may
have a simple pole at zero in the case
of a Scr\"odinger operator on the line.

2) Asymptotics \er{asDhl}
shows that the determinant $D$ for the operator
on the half-line has a pole of order $\le 1$ at zero.
Similarly, asymptotics \er{asDat0l}
shows that the determinant $D$ for the operator
on the line has a pole of order $\le 4$ at zero.

3) Coefficients of the asymptotics \er{asDhl} are
expressed in terms of auxiliary boundary
problems. It gives a complete description
of the determinant at zero in Case~1.
The asymptotics \er{Dasimp} in Case~2
is more complicated.
In this case we give necessary and sufficient conditions for the determinant
to have a pole of maximal order 4 at zero.

4) The asymptotics \er{asDhl}, \er{asDat0l} were used
for trace formulas in terms of resonances in the papers
\cite{BK16}, \cite{BK16x}.

\medskip

Jost \cite{J47} expressed the determinant for a Schr\"odinger
operator in terms of Wronskian of some specific solutions (Jost
solutions). This relation is important to study spectral properties
of the Schr\"odinger operator, inverse problems,  and, in particular,
resonances (zeros of the determinant). The resonances of
Schr\"odinger operators studied by Zworski \cite{Z87},  Froese
\cite{F97}, Hitrik \cite{H99},
Simon \cite{S00}, Korotyaev \cite{K04}, \cite{K05}.

Many papers are devoted to the
scattering for one dimensional higher order $\ge 3$ operators,
see the book \cite{BDT88} and references therein.
In connection with the Boussinesq equation
the spectral problems for third order operators on the line considered
by Deift, Tomei and Trubowitz \cite{DTT82}.
The corresponding periodic problem was studied by McKean \cite{McK81}.
Resonances for third order operators on the line were considered by
Korotyaev \cite{K16}.
The spectral problems for fourth order operators on the line considered
by Iwasaki \cite{Iw88}, \cite{Iw88x},
Hoppe, Laptev and \"Ostensson \cite{HLO06}.
Resonances for fourth order operators
considered by Badanin and Korotyaev  \cite{BK16}, \cite{BK16x}.
Fourth order operators with periodic coefficients
studied in \cite{BK11}.

The plan of the paper is as follows.
In Section 2 we consider the operator on the half-line.
The operator on the line is considered in
Section 3.

\section{Operator on the half-line}
\setcounter{equation}{0}

\subsection{Jost function}
We consider Case 1.
Define the sets
\[
\lb{setsS}
\mS=\Big\{k\in\C:\arg k\in\big(0,{\pi\/4}\big)\Big\},
\qqq \mS(r)=\Big\{k\in\mS:|k|>r\Big\},\qq r>0.
\]
Consider the equation on $\R_+$
\[
\lb{iehl}
y''''+2(py')'+qy=k^4 y,\qqq k\in\C,
\]
in the class of function $y$ satisfying the conditions
\[
\lb{achl}
y^{[j]}\in AC_{loc}(\R_+),\qq j=0,1,2,3,
\]
where $AC_{loc}(\R_+)$ is the class of locally absolutely continuous functions
on $\R_+$.
Let $\p_1(x,k)$, $\p_2(x,k)$, $(x,k)\in\R_+\ts\mS$, be Jost solutions of
equation \er{iehl} satisfying the conditions
\[
\lb{aspjx}
\p_1(x,k)=e^{-kx},\qq\p_2(x,k)=e^{ikx},\qq\as\qq x>\g,
\]
for all $k\in \mS.$
Each function $\p_j(x,\cdot),j=1,2,x\in\R_+$, is analytic
function of $k\in\mS(r)$ for some $r>0$ large
enough, see \cite{Iw88}.
Introduce the function $w(k)$ by
\[
\lb{Delt}
w(k)=\p_1(0,k)\p_2''(0,k)-\p_1''(0,k)\p_2(0,k),\qqq
k\in\mS(r).
\]

\begin{lemma}
\lb{LmDet}
Let $p,q\in\cH$. Then
the function $w(k)$ has an analytic extension from $\mS(r)$
onto the whole complex plane and satisfies
\[
\lb{wfs}
\begin{aligned}
w(k)=-ke^{(i-1)k\g}\Big((1+i)\D_{34 }(k)
+2k\D_{24 }(k)
+(1-i)k^2\big(\D_{14 }(k)+\D_{23 }(k)\big)
\\
-2ik^{3}\D_{13 }(k)
-(1+i)k^{4}\D_{12 }(k)
\Big),\qq \forall\qq k\in\C,
\end{aligned}
\]
where $\D_{j\ell}$ are given by \er{defD}.
In particular, $w(0)=0$.
\end{lemma}

\no {\bf Proof.}
Let $k\in\mS(r)$. We have the identities
\[
\lb{psi01}
\ma\p_j^{[0]}\\\p_j^{[1]}\\\p_j^{[2]}\\\p_j^{[3]}\am(0)
=M^{-1}\ma\p_j^{[0]}\\\p_j^{[1]}\\\p_j^{[2]}\\\p_j^{[3]}\am(\g),\qq
j=1,2,
\]
here and below $\p_j(x)=\p_j(x,k),...$
The matrix $M$ satisfies the identity
$$
M^{-1}=-JM^\top J,\qq \where\qq
J=(J_{j\ell})_{j,\ell=1}^4=\ma 0&0&0&1\\0&0&-1&0\\0&1&0&0\\-1&0&0&0\am.
$$
Identities $J_{j\ell}=(-1)^{j+1}\d_{j,5-\ell}$
yield
\[
\lb{MM-1}
(M^{-1})_{jn}=-\sum_{\ell,m=1}^4 J_{j\ell}M_{m\ell}J_{mn}
=-\sum_{\ell,m=1}^4(-1)^{j+m}\d_{j,5-\ell}\d_{m,5-n}M_{m\ell}
=(-1)^{j+n}M_{5-n,5-j}
\]
for all $j,n\in\N_4$.
Identities \er{psi01}, \er{MM-1} give
\[
\lb{pjpr}
\begin{aligned}
\p_j^{[m-1]}(0)
=\sum\limits_{n=1}^4(M^{-1})_{mn}\p_j^{[n-1]}(\g)
=\sum\limits_{n=1}^4(-1)^{m+n}M_{5-n,5-m}\p_j^{[n-1]}(\g)
\\
=\sum\limits_{\ell=1}^4(-1)^{m+\ell-1}M_{\ell,5-m}
\p_j^{[4-\ell]}(\g),
\end{aligned}
\]
for all $j=1,2,m\in\N_4.$
Using the identities
$\p_1(x)=e^{-kx},\p_2(x)=e^{ikx}$ as $ x> \g,$ and \er{achl}
we obtain
$$
\p_1^{[\ell-1]}(\g)=(-k)^{\ell-1}e^{-k\g},\qqq
\p_2^{[\ell-1]}(\g)=(ik)^{\ell-1}e^{ik\g},\qqq\ell\in\N_4.
$$
Then for $m\in\N_4$ the identities \er{pjpr} give
$$
\p_1^{[m-1]}(0)
=(-1)^{m-1}e^{-k\g}\sum\limits_{\ell=1}^4k^{4-\ell}M_{\ell,5-m},
\qq
\p_2^{[m-1]}(0)
=(-1)^{m-1}e^{ik\g}\sum\limits_{\ell=1}^4(-ik)^{4-\ell}M_{\ell,5-m}.
$$
Substituting these identities into \er{Delt} we obtain
$$
\begin{aligned}
w(k)=e^{(i-1)k\g}\det\ma
\sum\limits_{\ell=1}^4k^{4-\ell}M_{\ell 4}(k)&
\sum\limits_{\ell=1}^4(-ik)^{4-\ell}M_{\ell 4}(k)\\
\sum\limits_{j=1}^4k^{4-j}M_{j 2}(k)&
\sum\limits_{j=1}^4(-ik)^{4-j}M_{j 2}(k)\am
\\
=e^{(i-1)k\g}\sum_{j,\ell=1}^4(i^{j}-i^{\ell})k^{8-j-\ell}
M_{j2}(k)M_{\ell 4}(k)
=e^{(i-1)k\g}\sum_{1\le j<\ell\le 4}(i^{j}-i^{\ell})k^{8-j-\ell}
\D_{j\ell }(k),
\end{aligned}
$$
which yields
\er{wfs}.
\BBox

\subsection{Resolvents}
Equation \er{ie} may be rewritten in the vector form
\[
\lb{me}
{\bf y}'=Q{\bf y},\qqq \where\qq
{\bf y}=\ma y^{[0]}\\y^{[1]}\\y^{[2]}\\y^{[3]}\am,
\qqq Q=\ma
0&1&0&0\\0&0&1&0\\0&-2p&0&1\\k^4-q&0&0&0\am.
\]
Introduce the matrix-valued function
\[
\lb{lsA}
 X (x,k)=\ma\vp_2^{[0]}&\vp_4^{[0]}&\p_1^{[0]}&\p_2^{[0]}\\
\vp_2^{[1]}&\vp_4^{[1]}&\p_1^{[1]}&\p_2^{[1]}\\
\vp_2^{[2]}&\vp_4^{[2]}&\p_1^{[2]}&\p_2^{[2]}\\
\vp_2^{[3]}&\vp_4^{[3]}&\p_1^{[3]}&\p_2^{[3]}
\am(x,k),\qqq (x,k)\in\R_+\ts\mS(r).
\]
The function $ X$ satisfies  equation \er{me}.
The conditions \er{iecond} and the definition \er{Delt}
imply that the
function $\det X(x,\cdot)$ has an analytic extension
from $\mS(r)$ onto the whole complex plane and satisfies
\[
\lb{dPD}
\det X(x,k)=\det X(0,k)=-w(k)\qqq\forall\qq(x,k)\in\R_+\ts\C.
\]

Let $(x,k)\in\R_+\ts\mS(r)$ for some $r>0$ large enough.
Then $\det X(x,k)\ne 0$ and we
introduce the matrix-valued functions
\[
\lb{GP}
\G(x,k)=\big(\G_{j\ell}(x,k)\big)_{j,\ell=1}^4= X^{-1}(x,k),
\]
\[
\lb{matX}
\O(x,k)=\big(\O_{j\ell}(x,k)\big)_{j,\ell=1}^4=\G(x,k)\dot{ X}(x,k)
= X^{-1}(x,k)\dot{ X}(x,k),\] where $\dot f=\fr{\pa f}{\pa k}.$

\begin{lemma} Let $p,q\in \cH,(x,k)\in\R_+\ts\mS(r)$.
Then the matrix-valued function $\O(x,k)$ satisfies
\[
\lb{Ompr}
\O'=4k^3\ma \G_{14}\vp_2&\G_{14}\vp_4&\G_{14}\p_1&\G_{14}\p_2\\
\G_{24}\vp_2&\G_{24}\vp_4&\G_{24}\p_1&\G_{24}\p_2\\
\G_{34}\vp_2&\G_{34}\vp_4&\G_{34}\p_1&\G_{34}\p_2\\
\G_{44}\vp_2&\G_{44}\vp_4&\G_{44}\p_1&\G_{44}\p_2 \am.
\]
\end{lemma}

\no {\bf Proof.}
Differentiating the equation $ X'=Q X$ with respect to $k$ we
obtain
$
\dot{ X}'=Q\dot{ X}+\dot Q X.
$
Moreover, the definition \er{GP} gives
$$
\G'=( X^{-1})'=- X^{-1} X' X^{-1}=- X^{-1}Q=-\G Q.
$$
Then the definition \er{matX} yields
\[
\lb{idOmpr}
\O'=(\G\dot X)'=\G'\dot{ X}+\G\dot{ X}'=-\G Q\dot{ X}+\G(Q\dot{ X}
+\dot Q X) =\G\dot Q X.
\]
The definition \er{me} of $Q$ gives
$$
\dot Q=4k^3\ma
0&0&0&0\\0&0&0&0\\0&0&0&0\\1&0&0&0\am.
$$
Substituting this identity and \er{lsA}
into the identity \er{idOmpr} we obtain \er{Ompr}.
\BBox

\medskip

We need the following result.

\begin{lemma} Let $p,q\in \cH,k\in\mS(r)$. Then

i) The resolvent $(H-k^4)^{-1}$ is an integral
operator with the kernel $R(x,t,k), x,t\in\R_+$, given by
\[
\lb{prres}
R(x,t,k)=\ca -\G_{14}(t,k)\vp_2(x,k)-\G_{24}(t,k)\vp_4(x,k),\ x<t
\\
\ \ \G_{34}(t,k)\p_1(x,k)+\G_{44}(t,k)\p_2(x,k),\ x>t \ac .
\]

ii) The kernel $R(x,t,k)$ satisfies
\[
\lb{idresder}
R(x,x,k)=-\fr{F_{1}'(x,k)}{4k^3}=\fr{F_{2}'(x,k)}{4k^3}
\]
for all $x\in\R_+$,
where
\[
\lb{defF12}
F_1=\O_{11}+\O_{22},\qqq
F_2=\O_{33}+\O_{44}.
\]
Moreover,
\[
\lb{F10}
F_1(0,k)=0,\qqq
F_2(x,k)=F_2^0(x,k)\qq\as\qq x>\g,
\]
where $F_2^0$ is equal to $F_2$ for $p=q=0$.
\end{lemma}

\no {\bf Proof.}
i)
Let $(t,k)\in\R_+\ts\mS$.
The function $G(x)=R(x,t,k)$ satisfies
$$
\begin{aligned}
G''''+2(pG')'+qG=k^4 G,\qqq \forall\qq x\in\R_+\sm\{t\},
\\
G,G',G''\in AC_{loc}(\R_+),\qqq
G^{[3]}\in AC_{loc}(0,t)\cap AC_{loc}(t,+\iy),
\\
G^{[3]}(t+0)-G^{[3]}(t-0)=1,\qqq
G(0)=G''(0)=0,\qqq
\lim_{x\to\iy}G(x)=0.
\end{aligned}
$$
These conditions give
$$
R(x,t,k)=\ca -A_1(t,k)\vp_2(x,k)-A_2(t,k)\vp_4(x,k),\ x<t
\\
\ \ A_3(t,k)\p_1(x,k)+A_4(t,k)\p_2(x,k),\ x>t \ac,
$$
where $A_j(t,k),j\in\N_4$ satisfy
$$
 X(t,k)\ma A_1\\A_2\\ A_3\\A_4\am(t,k)=\ma0\\0\\0\\1\am.
$$
The definition \er{GP} gives
$A_j=\G_{j4},j\in\N_4$, which yields \er{prres}.

ii) The identities \er{Ompr} and  \er{prres} imply
\er{idresder}.

Let $x>\g$. Then $p=q=0$ and
$
\O(x,k)=\O^0(x,k),
$
where $\O^0(x,k)$ is equal to $\O(x,k)$ for $p=q=0$.
This yields the second identity in \er{F10}.

The identity \er{lsA} gives $\dot{ X}_{\ell j}(0,k)=0,\ell\in\N_4,j=1,2$.
The definitions \er{matX}, \er{defF12} imply
$F_1=(\G\dot X)_{11}+(\G\dot X)_{22}$.
Then
$$
F_1(0,k)
=\sum_{\ell=1}^4\big(\G_{1\ell}(0,k)\dot{ X}_{\ell 1}(0,k)
 +\G_{2\ell}(0,k)\dot{ X}_{\ell 2}(0,k)\big)=0,
$$
which yields the first identity in \er{F10}.
\BBox

\subsection{Determinant}
Now we express the determinant $D(k)$ in terms of the Jost function
$w(k)$.

\begin{lemma}
\lb{Dd}
Let $p,q\in\cH$. Then
the determinant $D$, given by \er{a.2}, satisfies
\[
\lb{idDk}
D(k)=-\fr{w(k)}{2k^2}\qqq\forall\qq k\in\C\sm\{0\},
\]
where $w$ is given by \er{Delt}. Moreover,
\[
\lb{idDkhl}
D=\fr{e^{(i-1)k\g}}{2k}\Big((1+i)\D_{34 }
+2k\D_{24 }
+(1-i)k^2\big(\D_{14 }+\D_{23 }\big)
-2ik^{3}\D_{13 }
-(1+i)k^{4}\D_{12 }
\Big)
\]
on $\C\sm\{0\}$, where $\D_{j\ell}$ are given by \er{defD}.
\end{lemma}

\no {\bf Proof.}
Let $0<x<\g<\a<\iy$ and let $k\in\mS(r)$.
Integrating the first identity
in \er{idresder} in the
interval $(0,x)$ and the second identity
in \er{idresder} in the interval $(x,\a)$
and using \er{defF12} and \er{F10}
we obtain
\[
\lb{irx}
\int_{0}^{\a}R(t,t,k)dt=
\fr{1}{4k^3}\Big(\int_{x}^{\a}F_{2}'(t,k)dt
-\int_{0}^{x}F_{1}'(t,k)dt \Big)
=-\fr{1}{4k^3}\big(\Tr \O(x,k)-F_2^0(\a,k)\big).
\]
Furthermore, identities \er{dPD} and \er{GP} imply
\[
\lb{TrX} \Tr \O=\Tr \G\dot{ X}=\Tr
 X^{-1}\dot{ X}=\fr{1}{\det X}\fr{\pa\det X}{\pa k}
=\fr{\dot w}{w}.
\]
Substituting \er{TrX} into \er{irx} we obtain
$$
\int_{0}^{\a}R(t,t,k)dt
=-\fr{1}{4k^3}\Big(\fr{\dot w(k)}{w(k)}-F_2^0(\a,k)\Big),
$$
Using a similar identity for $p=q=0$ we obtain
$$
\Tr\big(R(k)-R_0(k)\big)=
\int_{0}^{\a}\big(R(t,t,k)-R_0(t,t,k)\big)dt
=-\fr{1}{4k^3}\Big(\fr{\dot{w}(k)}{w(k)}
-\fr{\dot{w_0}(k)}{w_0(k)}\Big),
$$
where $R_0,w_0$ are equal to $R,w$ as $p=q=0$ respectively.
The identity
$$
{D'(k)\/D(k)}=\Tr R_0(k)VR(k)=\Tr (R_0(k)-R(k)),
$$
yields
$$
D(k)=\fr{w(k)}{w_0(k)}.
$$
The identity $w_0(k)=-2k^2$ gives \er{idDk}.
The identities \er{wfs} and \er{idDk} imply
\er{idDkhl}.
$\BBox$

\medskip

We are ready to prove Theorem \ref{T2}.

\medskip

\no {\bf Proof of Theorem \ref{T2}.}
The identity
\er{idDkhl} yields the asymptotics \er{asDhl}.

We prove \er{detne0}.
 Assume that
$\D_{34}(0)=\D_{24}(0)=\D_{14}(0)+\D_{23}(0)=0.$
Then the definitions \er{defD} give
$$
\det\ma M_{3 2}&M_{3 4}\\M_{4 2}&M_{4 4}\am(0)
=\det\ma M_{2 2}&M_{2 4}\\M_{4 2}&M_{4 4}\am(0)=
\det\ma M_{1 2}&M_{1 4}\\M_{4 2}&M_{4 4}\am(0)
+\det\ma M_{2 2}&M_{2 4}\\M_{3 2}&M_{3 4}\am(0)=0.
$$
These identities imply
$$
\det\ma M_{2 2}&M_{2 4}\\M_{3 2}&M_{3 4}\am(0)=
\det\ma M_{1 2}&M_{1 4}\\M_{4 2}&M_{4 4}\am(0)=0.
$$
Then
$\det M(0)=0$. It is in contradiction with
the Liouville identity $\det M=1$, which proves \er{detne0}.

(i) The asymptotics \er{asDhl}
gives that the function $D$ is entire iff
$ \D_{34}(0)=0$.  The first identity in \er{sps} shows
that in this case $\l=0$ is an eigenvalue of the operator
$H_1$.

(ii) The asymptotics \er{asDhl} gives that the function $k^{-1}D$ is entire iff
$ \D_{34}(0)=\D_{24}(0)=0$. The first and second identities in \er{sps}
show that in this case $\l=0$ is an eigenvalue of the operators
$H_1$ and $H_2$.
\BBox



\subsection{Square of a second order operator.}
Now we consider the square
of a second order operator.

\begin{proposition}
\lb{PropSqSchr}
Let $p,p''\in L^1(\R_+)$ and $\supp p\in[0,\g]$. Let $H=h^2$, where
$h$ is the operator $h=-\pa^2-p$ on $\R_+$
with the condition $y(0)=0$.
Then
the determinant $D$ satisfies the identity
\[
\lb{idDh^2}
D(k)=\p(0,ik)\p(0, k)\qqq \forall\qq k\in\C,
\]
where
$\p(x,k),(x,k)\in\R_+\ts\C$, is the solution
of the equation
$
-y''-py=z y,
$
satisfying the identity
\[
\lb{defpsi}
\begin{aligned}
\p(x,k)=e^{ikx}\qqq\forall\qq x>\g,\\
k=\sqrt z,\qq
k\in\C_+\qq \as \qq z\in\C\sm\R_+.
\end{aligned}
\]

\end{proposition}

\no {\bf Proof.}
In the considered case we have
$
\p_1(x,k)=\p(x,ik),\p_2(x,k)=\p(x,k).
$
Using the identities $\p''(x,k)=(-p(x)- k^2)\p(x,k)$
we obtain
$$
\begin{aligned}
w(k)=\p_1(0,k)\p_2''(0,k)-\p_1''(0,k)\p_2(0,k)
=\p(0,ik)\p''(0,k)-\p(0,k)\p''(0,ik)
\\
=\p(0,ik)(-p(0)- k^2)\p(0, k)-\p(0,k)(-p(0)+ k^2)\p(0,ik)
=-2k^2\p(0,ik)\p(0, k).
\end{aligned}
$$
The identity \er{idDk} yields \er{idDh^2}.
\BBox

\medskip

\no {\bf Remark.} The identity \er{idDh^2} used in \cite{BK16} in order to
describe resonances of $h^2$.

\section{Operator on the line}
\setcounter{equation}{0}

\subsection{Jost function} We consider Case 2.
Let $\mS,\mS(r),r>0$ be the sets given
by the relations \er{setsS}.
Consider the equation on $\R$
\[
\lb{iel}
y''''+2(py')'+qy=k^4 y,\qqq k\in\C,
\]
in the class of functions $y$ satisfying the conditions
$$
y^{[j]}\in AC_{loc}(\R),\qq j=0,1,2,3.
$$
Let $\p_j(x,k),j\in\N_4$, be Jost solutions of
equation \er{iel} satisfying the conditions
\[
\lb{aspjxl}
\begin{aligned}
\p_1(x,k)=e^{-kx},\qq\p_2(x,k)=e^{ikx},\qq\as\qq x>\g,\\
\p_3(x,k)=e^{-ikx},\qq\p_4(x,k)=e^{kx},\qq\as\qq x<0,
\end{aligned}
\]
for all $k\in \mS.$ Each function
$\p_j(x,\cdot),(j,x)\in\N_4\ts\R$, is analytic in
$k\in\mS(r)$ for some $r>0$ large enough, see \cite{Iw88}.

Let $k\in\mS(r).$
Introduce the Wronskian $w(k)$ by
\[
\lb{Deltl}
w(k)=\det\ma\p_1^{[0]}&\p_2^{[0]}&\p_3^{[0]}&\p_4^{[0]}\\
\p_1^{[1]}&\p_2^{[1]}&\p_3^{[1]}&\p_4^{[1]}\\
\p_1^{[2]}&\p_2^{[2]}&\p_3^{[2]}&\p_4^{[2]}\\
\p_1^{[3]}&\p_2^{[3]}&\p_3^{[3]}&\p_4^{[3]}
\am(x,k),\qq k\in\mS(r).
\]
\"Ostensson and Yafaev \cite{OY12} proved that the function $w$
has an analytic extension from $\mS(r)$ onto the whole
complex plane and
\[
\lb{OYf}
D(k)=-\fr{w(k)}{16ik^6}\qqq\forall\qq k\in\C\sm\{0\}.
\]
This formula shows
that the determinant may have a pole of order $\le 6$ at zero.
We prove that the order is $\le 4$.

Introduce the functions $\D_{jn,\ell m}(k),1\le j<n\le 4,
1\le\ell<m\le 4,k\in\C$ by
$$
\D_{jn,\ell m}=\det\ma M_{j\ell}& M_{jm}\\M_{n\ell}& M_{nm}\am.
$$
Each function $\D_{jn,\ell m}$ is an entire function of the
variable $\l=k^4$.

\begin{lemma}
Let $p,q\in \cH$.
Then
the determinant $D(k)$ satisfies
\[
\lb{Dline}
D(k)=-\fr{e^{(i-1)k\g}}{8k^4}\sum_{n=0}^8A_n(k)k^{n},\qqq k\in\C\sm\{0\},
\]
where $A_n(k)$ are entire in $\l=k^4$ functions given by
\[
\lb{dAn}
\begin{aligned}
A_0=\D_{34,12},\qqq A_1=(1-i)(\D_{34,13}-\D_{24,12}),
\\
A_2=i(2\D_{24,13}-\D_{23,12}-\D_{14,12}-\D_{34,14}-\D_{34,23}),
\\
A_3=(1+i)(\D_{13,12}-\D_{23,13}-\D_{14,13}+\D_{24,14}+\D_{24,23}-\D_{34,24}),\\
A_4=2\D_{24,24}-\D_{12,12}+2\D_{13,13}-
\D_{23,14}-\D_{14,14}-\D_{23,23}-\D_{14,23}-\D_{34,34},
\\
A_5=(1-i)(\D_{24,34}-\D_{12,13}+\D_{13,14}+\D_{13,23}-\D_{23,24}-\D_{14,24}),
\\
A_6=i(\D_{12,14}+\D_{12,23}+\D_{23,34}+\D_{14,34}-2\D_{13,24}),
\\
A_7=(1+i)(\D_{12,24}-\D_{13,34}),\qqq
A_8=\D_{12,34}.
\end{aligned}
\]
Moreover,
\[
\lb{Dasimp}
D(k)=-\fr{e^{(i-1)k\g}}{8k^4}\Big(A_0(0)+A_1(0)k+A_2(0)k^2+A_3(0)k^3
+O(k^4)\Big),
\]
as $k\to 0$ uniformly in $\arg k\in[0,2\pi]$.
\end{lemma}

\no {\bf Proof.} Substituting the identities
$$
\ma\p_j^{[0]}\\\p_j^{[1]}\\\p_j^{[2]}\\\p_j^{[3]}\am(\g,k)
=M(k)\ma\p_j^{[0]}\\\p_j^{[1]}\\\p_j^{[2]}\\\p_j^{[3]}\am(0,k),\qqq
j=3,4,
$$
into the definition \er{Deltl} we obtain
\[
\lb{wl}
w(k)=\sum_{\ell,m=1}^4W_{\ell
m}(k)\p_3^{[\ell-1]}(0,k)\p_4^{[m-1]}(0,k) =\sum_{1\le\ell<m\le
4}W_{\ell m}(k)E_{\ell m}(k),
\]
where
\[
\lb{Elmpr}
E_{\ell m}(k)=\det\ma\p_3^{[\ell-1]}&\p_3^{[m-1]}
\\\p_4^{[\ell-1]}&\p_4^{[m-1]}\am (0,k),
\]
\[
\lb{Wlmpr}
W_{\ell m}(k)=\det\ma\p_1^{[0]}(\g,k)&\p_2^{[0]}(\g,k)&M_{1\ell}(k)&M_{1m}(k)\\
\p_1^{[1]}(\g,k)&\p_2^{[1]}(\g,k)&M_{2\ell}(k)&M_{2m}(k)\\
\p_1^{[2]}(\g,k)&\p_2^{[2]}(\g,k)&M_{3\ell}(k)&M_{3m}(k)\\
\p_1^{[3]}(\g,k)&\p_2^{[3]}(\g,k)&M_{4\ell}(k)&M_{4m}(k)
\am.
\]
Recall that $\p_j^{[\ell-1]}(\cdot,k)\in AC_{loc}(\R)$ for all
$j,\ell\in\N_4$.
Then the definitions \er{aspjxl} imply
\[
\lb{pjl}
\p_1^{[j]}(\g,k)=(-k)^je^{-k\g},
\qq\p_2^{[j]}(\g,k)=(ik)^{j}e^{ik\g},
\qq
\p_3^{[j]}(0,k)=(-ik)^{j},
\qq
\p_4^{[j]}(0,k)=k^{j},
\]
for all $j\in\N_4.$
Substituting the identities \er{pjl} into \er{Elmpr}
we have
$$
E_{\ell m}=\big((-i)^{\ell-1}-(-i)^{m-1}\big)k^{\ell+m-2},
$$
The identity \er{wl}
gives
\[
\lb{wl2}
w=
(1+i)kW_{12}
+2k^{2}W_{1 3}
+(1-i)k^{3}\big(W_{1 4}+W_{2 3}\big)
-2ik^{4}W_{2 4}
-(1+i)k^{5}W_{3 4}.
\]
Substituting the identities \er{pjl} into
\er{Wlmpr} we obtain
$$
\begin{aligned}
W_{\ell m}=e^{(i-1)k\g}\det\ma1&1&M_{1\ell}&M_{1m}\\
-k&ik&M_{2\ell}& M_{2m}\\
k^2&-k^2& M_{3\ell}& M_{3m}\\
-k^3&-ik^3& M_{4\ell}& M_{4m}
\am
=e^{(i-1)k\g}\Big((1+i)k\D_{34,\ell m}-2k^{2}\D_{24,\ell m}
\\
+(1-i)k^{3}(\D_{23,\ell m}+\D_{14,\ell m})
+2ik^{4}\D_{13,\ell m}-(1+i)k^{5}\D_{12,\ell m}\Big).
\end{aligned}
$$
Substituting this identity into \er{wl2} we obtain
$$
w=2ik^2e^{(i-1)k\g}\sum_{n=0}^8A_nk^{n},
$$
where $A_n$ are given by \er{dAn}.
The identity \er{OYf} gives \er{Dline},
which yields the asymptotics \er{Dasimp}.
\BBox

\medskip

\no {\bf Proof of Theorem \ref{ThDentl}.}
The asymptotics \er{Dasimp} gives \er{asDat0l}.
The asymptotics \er{asDat0l} implies
that the function $D$ has a pole of order $\le 3$
at $k=0$ iff $\F(0)=0.$
The identity \er{sps} shows that this is
so iff $\l=0$ is an eigenvalue of the
operator $H_3$.
\BBox

\subsection{Square of a second order operator.}
Now we consider the square
of a second order operator.

\begin{proposition}
\lb{PropSqSchrl}
Let $p,p''\in L^1(\R)$ and $\supp p\in[0,\g]$. Let $H=h^2$, where
$h$ is the operator $h=-\pa^2-p$ on $\R$.
Then the determinant $D$ satisfies the identity
\[
\lb{idDh^2l}
D(k)=\fr{i}{4k^2}\{\p_+(x,k),\p_-(x,k)\}
\{\p_+(x,ik),\p_-(x,ik)\}\qqq \forall\qq k\in\C\sm\{0\},
\]
where
$
\{f,g\}=fg'-f'g
$ and
$\p_\pm(x,k),(x,k)\in\R\ts\C$, are solutions
of the equation
$
-y''-py=z y,
$
satisfying the identity
\[
\lb{defpsil}
\begin{aligned}
\p_+(x,k)=e^{ikx}\qq\forall\qq x>\g,\qqq
\p_-(x,k)=e^{-ikx}\qq\forall\qq x<0,\\
k=\sqrt z,\qq
k\in\C_+\qq \as \qq z\in\C\sm\R_+.
\end{aligned}
\]
\end{proposition}

\no {\bf Proof.}
 In the considered case we have
$$
\p_1(x,k)=\p_+(x,ik),\qq\p_2(x,k)=\p_+(x,k),
\qq\p_3(x,k)=\p_-(x,k),\qq\p_4(x,k)=\p_-(x,ik).
$$
Substituting the identities $\p_\pm''(x,k)=(-p(x)- k^2)\p_\pm(x,k)$
into \er{Deltl} we obtain
$$
w(k)=k^4\det\ma\p_+(x,ik)&\p_+(x,k)&\p_-(x,k)&\p_-(x,ik)\\
\p_+'(x,ik)&\p_+'(x,k)&\p_-'(x,k)&\p_-'(x,ik)\\
\p_+(x,ik)&-\p_+(x,k)&-\p_-(x,k)&\p_-(x,ik)\\
\p_+'(x,ik)&-\p_+'(x,k)&-\p_-'(x,k)&\p_-'(x,ik)\am,
$$
which yields
$$
w(k)=4k^4\{\p_+(x,k),\p_-(x,k)\}
\{\p_+(x,ik),\p_-(x,ik)\}.
$$
The identity \er{OYf} gives \er{idDh^2l}.
\BBox

\medskip

\no {\bf Remark.} The identity \er{idDh^2l} used in \cite{BK16x} in order to
describe resonances of $h^2$.

\medskip

\no {\bf Acknowledgments.}
\small The study was supported
by the RNF grant  No 15-11-30007.


\begin{thebibliography}
{999}\setlength{\itemsep}{-\parskip} \footnotesize

\bibitem[BK16]{BK16} Badanin, A., Korotyaev, E.
Resonances for Euler-Bernoulli operator on the half-line.
Preprint, 2016.

\bibitem[BK16x]{BK16x} Badanin, A., Korotyaev, E.
Resonances of 4-th Order Differential Operators.
Preprint, 2016.

\bibitem[BK11]{BK11} Badanin, A.,  Korotyaev, E.
Spectral estimates for periodic fourth order operators.
St.Petersburg Math. J. 22:5 (2011) 703--736.

\bibitem[BDT88]{BDT88} Beals, R., Deift, P., Tomei, C.
Direct and inverse scattering on the line, Mathematical survays and
monograph series, No. 28, AMS, Providence, 1988.

\bibitem[DTT82]{DTT82} Deift, P., Tomei, C., Trubowitz, E.
Inverse scattering and the Boussinesq equation,
Comm. on Pure and Appl. Math. 35 (1982) 567--628.

\bibitem[F97]{F97} Froese, R. Asymptotic distribution
of resonances in one dimension. Journal of differential equations,
137 (1997), 251--272.

\bibitem[H99]{H99} Hitrik, M. Bounds on scattering poles in one dimension.
Comm. Math. Phys. 208 (1999), no. 2, 381--411.

\bibitem[HLO06]{HLO06} Hoppe, J., Laptev, A., \"Ostensson, J.
Solitons and the removal of eigenvalues for fourth-order differential
operators,  Int. Math. Res. Not., (2006), 14 pp.

\bibitem[Iw88]{Iw88} Iwasaki, K.
Scattering Theory for 4-th Order Differential Operators, I.
Japanese journal of mathematics. New series, 14(1) (1988), 1--57.

\bibitem[Iw88x]{Iw88x} Iwasaki, K.
 Scattering theory for 4-th order differential operator, II.
Japanese journal of mathematics.
New series, 14(1) (1988), 59--96.

\bibitem[J47]{J47} Jost, R. Uber die falschen Nullstellen der Eigenwerte der S-Matrix.
Helvetica Physica Acta, 20(3) (1947), 256--266.

\bibitem[K04]{K04}   Korotyaev, E. Inverse resonance scattering on
the half line.  Asymptot. Anal. 37 (2004), no. 3-4, 215--226.

\bibitem[K05]{K05}   Korotyaev, E. Inverse resonance scattering on
the real line. Inverse Problems 21.1 (2005),  325--341.

\bibitem[K16]{K16}
Korotyaev, E. Resonances  of third order differential operators.
ArXiv preprint arXiv:1605.01842 (2016).

\bibitem[McK81]{McK81} McKean, H. Boussinesq's equation on the circle,
Com. Pure and Appl. Math. 34 (1981) 599--691.


\bibitem[OY12]{OY12} \"Ostensson, J., Yafaev, D. R.
A trace formula for differential operators of arbitrary order.
In A panorama of modern operator theory and related topics. Springer Basel, 2012,
541--570.

\bibitem[S00]{S00} Simon, B. Resonances in one dimension and Fredholm
 determinants,  J. Funct. Anal. 178 (2000), no. 2, 396--420.

 \bibitem[Z87]{Z87} Zworski, M. Distribution of poles for scattering
on the real line, J. Funct. Anal. 73(1987), 277--296.


\end{thebibliography}
\end{document}